\documentclass[aip,cha,amsmath,amssymb,reprint]{revtex4-1}

\usepackage{graphicx}
\usepackage[utf8]{inputenc}
\usepackage[T1]{fontenc}
\usepackage{mathptmx}
\usepackage{etoolbox}

\usepackage{xcolor}
\usepackage{graphicx}
\usepackage{amsmath,amsfonts,amsthm,bm}
\usepackage{braket}
\renewcommand{\vec}[1]{\mathbf{#1}}

\begin{document}

%\title{Minimal Quantum Reservoirs with Hamiltonian Encoding}
\title{Minimal Quantum Reservoirs with Hamiltonian Encoding}

\author{Gerard McCaul}
\email[E-mail: ]{g.mccaul@lboro.ac.uk}
\affiliation{Department of Physics, Loughborough University, Loughborough LE11 3TU, UK}
\author{Juan Sebastian Totero Gongora}
\email[E-mail: ]{j.totero-gongora@lboro.ac.uk}
\affiliation{Emergent Photonics Research Centre, Department of Physics, Loughborough University, LE11 3TU, Loughborough, United Kingdom}
\author{Wendy Otieno}
\affiliation{Department of Physics, Loughborough University, Loughborough LE11 3TU, UK}
\author{Sergey Savel'ev}
\affiliation{Department of Physics, Loughborough University, Loughborough LE11 3TU, UK}
\author{Alexandre Zagoskin}
\affiliation{Department of Physics, Loughborough University, Loughborough LE11 3TU, UK}
\author{Alexander~G.~Balanov}
\email[E-mail: ]{a.balanov@lboro.ac.uk}
\affiliation{Department of Physics, Loughborough University, Loughborough LE11 3TU, UK}

%\author{List of Authors}
%\affiliation{Department of Physics, Loughborough University, LE11 3TU, Loughborough, United Kingdom}
%\affiliation{Emergent Photonics Research Centre, Department of Physics, Loughborough University, LE11 3TU, Loughborough, United Kingdom}

\date{\today}

\begin{abstract}
We investigate a minimal architecture for quantum reservoir computing based on Hamiltonian encoding, in which input data is injected via modulation of system parameters rather than state preparation. This approach circumvents many of the experimental overheads typically associated with quantum machine learning, enabling computation without feedback, memory, or state tomography. We demonstrate that such a minimal quantum reservoir, despite lacking intrinsic memory,  can perform nonlinear regression and prediction tasks when augmented with post-processing delay embeddings. 
%We demonstrate that a minimal memoryless quantum reservoir can perform nonlinear regression and prediction tasks when augmented with post-processing delay embeddings. 
Our results provide a conceptually and practically streamlined framework for quantum information processing, offering a clear baseline for future implementations on near-term quantum hardware.

% We introduce a minimal architecture for quantum reservoir computing based on Hamiltonian encoding, in which input data is injected via modulation of system parameters rather than state preparation. This approach circumvents many of the experimental overheads typically associated with quantum machine learning, enabling computation without feedback, memory, or state tomography. We demonstrate that a minimal memoryless quantum reservoir can perform nonlinear regression and prediction tasks when augmented with post-processing delay embeddings. Our results provide a conceptually and practically streamlined framework for quantum information processing, offering a clear baseline for future implementations on near-term quantum hardware.
% 

\end{abstract}

%\pacs{12345}% insert suggested PACS numbers in braces on next line

\maketitle %\maketitle must follow title, authors, abstract and \pacs

% {
% \color{red}
% \section*{Figures}
% \begin{itemize}
%     \item[Fig 1] (a) System Configuration (qubits, external field, main variables. (b) Encoding and readout scheme
%     \item[Fig 2] Post-processing delay engineering (Ludge-like, Juan). 
%     \item[Fig 3] Regression example: 3 cases: a) QR1 works for regression, b) QR1 doesn't work with another function, c) a new QR (QR2) works for the second function by changing parameters (or adding delays).  
%     \item[Fig 4] Regression massive span (Wendy). 
%     \item[Fig 5] RandomSin task (Juan).    
%     \item[Fig 7] Autonomous prediction: Random Sine waves (Juan).     
% \end{itemize}
% }
 {\bf Reservoir computing presents an efficient framework for harnessing the dynamics of complex physical systems to perform information processing tasks. By driving a high-dimensional dynamical system, acting as the “reservoir,” with an input signal and subsequently extracting outputs through simple, such as linear, operations on the system’s response, this paradigm facilitates computation through physical evolution itself. In contrast to conventional approaches to information processing, which typically rely on precise control of the signals involved, reservoir computing capitalises on the inherent richness of physical systems, transforming environmental noise and imperfections into computational assets. This approach holds significant promise in the quantum domain, where a high-dimensional state space can be achieved with a greatly reduced number of elements, and noise is present even at cryogenic temperatures. Recent advances have demonstrated the feasibility of quantum reservoir realisations based on quantum coherent systems, encompassing nonlinear electronic, spintronic, or photonic oscillators, as well as single atoms. However, a central challenge persists: how to effectively encode and extract information to exploit the unique properties of quantum dynamics. In this work, we explore an encoding strategy that embeds the input data directly into the Hamiltonian parameters governing the dynamics of the reservoir. This Hamiltonian encoding approach enables highly nonlinear feature mapping without the constraints of precise state preparation, providing a minimal yet potent route to quantum reservoir computing.
}

\section{Introduction} 

% \begin{quotation}
%     First paragraph is special on Chaos \cite{andreevEmergenceControlComplex2021}
% \end{quotation}

% \begin{itemize}
%     \item Introduction to QRC, recent results
%     \item State-of-the-art: state encoding, opportunities and challenges
%     \item Hamiltonian encoding: previous literature, advantages, challenges
% \end{itemize}

% \begin{itemize}
%     \item Hamiltonian encoding, experiment-driven methodology
%     \item Memory-less operation
%     \item Quick summary of results: regression and prediction
%     \item 1-2 lines on theoretical interpretation. 
% \end{itemize}

Reservoir computing \cite{Jaeger2004,Marcucci2020,Schuman2017,Verstraeten2007} has recently emerged as a powerful paradigm for information processing. In this framework, a complex, high-dimensional dynamical system—the reservoir—is driven by an input signal. The reservoir's intrinsic dynamics map this signal into a rich feature space, from which the output of a computation can be extracted via linear operations on the system’s response. While the idea of treating physical systems as information processors has long been used to derive physical limits from informational principles \cite{dariano_quantum_2016,lieb_finite_1972, LLoyd2014,PRXQuantum.2.010101}, quantum engineering \cite{zagoskin_quantum_2011, smirnov_modelling_2007, zagoskin_modeling_2007} - and reservoir computing in particular - renders this interpretation literal \cite{SethLloyd2000,Lloyd2002}.

The appeal of reservoir computing lies in its insensitivity to the specific dynamics chosen for the reservoir. This rests on the observation that any input-output relationship can be approximately linearised \cite{anco_invertible_2008, appeltant_information_2011} if the input is embedded in a high-dimensional space composed of linearly independent functions of that input. The specific character of those functions is irrelevant provided they span a sufficiently expressive basis. This paradigm makes a virtue of the apparent 'messiness' encountered in real systems: environmental noise and experimental imprecision — which degrade both classical and quantum algorithms — can \textit{enrich} the reservoir dynamics. Any physical system whose dynamics can nonlinearly embed input data into a higher-dimensional space can, in principle, function as a reservoir \cite{Yan:2024aa}. There is therefore a double attraction to quantum reservoirs: the exponential scaling of Hilbert space and the intrinsic presence of noise. Consequently, there has been a surge of interest in quantum reservoir computers \cite{Fujii2017, Ghosh2019, polaritoniccomputing}, explored in diverse platforms including both linear \cite{nokkala_gaussian_2021, nokkala_online_2023} and nonlinear oscillators \cite{Govia2020, angelatos_reservoir_2021}, as well as single atoms \cite{mccaul_towards_2023} and qudits \cite{Kalfus2021}.

The essential question for \textit{quantum} reservoir computing is how its underlying dynamics - and its relationship to measurement - both distinguish it and enhance its capacities relative to the classical case. There are, of course, many possible approaches to this problem; critically however, the performance of a quantum (or indeed any) reservoir computer hinges on the interplay between two essential features. First, the underlying complexity of the dynamics, set by the physical system itself. And second, the manner in which input is encoded and output is extracted. It is this mask of interpretation that transforms physical evolution into computation, and the question of how best to design this mask remains open.
% {\color{red} There are numerous potential approaches to investigating this problem. However, critically, the performance of a quantum (or indeed any) system as a reservoir computer hinges on the interplay between two fundamental features.
% Firstly, as previously discussed, the underlying complexity of the dynamics, determined by the system under consideration, plays a crucial role. Secondly, the method employed to both encode and read out information into this system is equally important.
% It is this intricate layer of interpretation that transforms physical evolution into computation, and the optimal approach to achieving this remains an open question. 
% }
%There are of course a plethora of potential approaches to investigating this problem, but critically the performance of a quantum (or indeed any) system as a reservoir computer will lie in the interplay between two essential features. 
%First, as outlined above, the underlying complexity of the dynamics, which is set by the system under consideration. Married to this however is how one chooses to both \textit{encode} and \textit{readout} information into this system.
% It is this mask of interpretation that transforms physical evolution into computation, and the question of how best to achieve this remains open. 

In many implementations of quantum reservoir computing (QRC), input data is encoded directly into the quantum state of the reservoir \cite{nokkala_gaussian_2021, li_estimating_2024, Palacios:2024aa}. This approach enables explicit use of quantum features such as entanglement, but at the cost of requiring arbitrary and high-fidelity state preparation. To encode information this way, each input must be mapped to a well-defined quantum state, which must then be reliably prepared and preserved. As a result, the familiar implementation challenges of quantum computing —decoherence, imprecision, and control overhead — are inherited wholesale by QRC in this regime. A natural alternative is therefore not to encode information into quantum states, but into the Hamiltonian parameters governing the reservoir’s evolution. 

In this work, we explore how such \emph{Hamiltonian encoding} can be used to drive a quantum system with input data, eliminating the need for state preparation.  We show that even without intrinsic memory, this scheme can perform nonlinear regression and prediction tasks when paired with post-processing delay embeddings \cite{xia_reservoir_2022}. We also compare Hamiltonian encoding to conventional state-based schemes and identify its structural advantages in terms of nonlinearity and experimental tractability. 

We begin by introducing the physical model and encoding protocol, which together form what we term a \textit{minimal} reservoir. By minimal, we mean a quantum reservoir with no intrinsic memory, no feedback or recurrent dynamics, no state preparation, and no entanglement requirements — relying solely on observable expectations from an evolution reset for each input. Sec. \ref{sec:results} presents numerical demonstrations of this minimal reservoir's performance on synthesis and prediction tasks, including stable forecasts using only a small number of qubits. We conclude in Sec. \ref{sec:discussion} with a discussion of the implications and practical scope of this minimal model.

% To that end, we investigate here the performance of such a scheme for a prototypical quantum system, and the degree to which its performance can be ascribed to genuinely quantum effects. Despite this austerity, we show that the reservoir dynamics of even a simple system are sufficiently nonlinear and expressive, this model remains capable of prediction tasks by effectively learning an \textit{autoencoding} of the input sequence.
% ---

% To do so, the remainder of the paper is organised as follows. In Sec. \ref{sec:reservoir} we outline the underlying structure of reservoir computing, and the idea that the underlying capacity of a reservoir can be characterised by the effective dimension spanned by its readouts. In Sec. \ref{sec:model} we introduce and motivate the candidate model under investigation, together with its factorised approximation. The comparative performance of these two models can then be used to infer the degree to which dynamical entanglement plays a role in computing capacity. The results of this investigation are presented in Sec. \ref{sec:results}, and we conclude with a discussion of these results in Sec. \ref{sec:discussion}.  

\section{A Minimal Quantum Reservoir \label{sec:Methods}} 
\subsection*{Reservoir Computing}
Any computation can be viewed as a mapping from a sequence of inputs \( \vec{x}_i \) to a sequence of outputs \( \vec{y}_j \). Formally, we may say that the $n$th output $\vec{y}_n$ corresponds to some function over the history of inputs up to the $n$th term. That is, 
\begin{equation}
    \vec{y}_n= f(\vec{x}_0, ..., \vec{x}_n).
\end{equation}
In general, this function may be highly nonlinear and depend on the entire input history. It need not preserve input dimensionality, and direct representations of it are typically intractable. A computation does, however, admit a closed-form description when its input-output map is linear:
\begin{equation}
    \vec{y} = A \vec{x} + \vec{b},
\end{equation}
where \( A \) is an \( N \times M \) matrix, and \( \vec{x} \in \mathbb{R}^M \), \( \vec{y}, \vec{b} \in \mathbb{R}^N \). A restriction to linearity may appear limiting, but it parallels the necessary conditions for \textit{physical} dynamics to be analytically tractable. This shared structure is more than analogy—computation and physical evolution are interchangeable: each can be recast in the language of the other.

The key insight that enables generalisation of both physical and computational dynamics is that \textit{linearity is a matter of dimensionality}. This idea underpins many modern machine learning methods, including Dynamic Mode Decomposition (DMD) \cite{kutz_dynamic_2016}. Although a system’s behaviour may appear nonlinear in its native (often low-dimensional) space, it may be lifted to a higher-dimensional feature space in which its evolution is approximately linear. To illustrate this concept, consider a dynamical system governed by a nonlinear differential equation:
\begin{equation}
    \frac{d\vec{y}(t)}{dt} = f(\vec{x}(t),t),
    \label{eq:diffeqexample}
\end{equation}
where $ f $ is a nonlinear function and $ \vec{y}(t) $ is the output. Although the mapping from inputs $ \vec{x}(t) $ to outputs $ \vec{y}(t) $ may be highly nonlinear, there exists a transformation $ \phi: \mathbb{R}^{M} \rightarrow \mathbb{R}^{K} $ (with $ K \gg M $) such that the dynamics of the lifted state evolve approximately linearly:
\begin{equation}
    \frac{d\phi(\vec{x}(t))}{dt} = A\, \phi(\vec{x}(t)),
\end{equation}
where $ A $ is a $ K \times K $ matrix. This construction reframes the original nonlinear mapping $ f(\vec{x}(t), t) $ as the projection of a lifted linear evolution: rather than computing Eq. \eqref{eq:diffeqexample} directly, one evolves a transformed state $ \phi(\vec{x}(t)) $ under a linear operator $ A $, and recovers the output as $ \vec{y}(t) \approx P\, \phi(\vec{x}(t)) $ for some projection $ P $. This perspective ultimately stems from Koopman operator theory \cite{PhysRev.40.749, Baker1958,Curtright2014,Groenewold1946, Koopman315,Wilkie1997a,Wilkie1997b, bondar2012,Bondar2013, PhysRevE.99.062121, mccaul_wave_2023, Koopman-nonlinear}, and allows one to interpret both physical and computational dynamics as a linear evolution in a high-dimensional \textit{feature} space, whose effective dimensionality governs the model’s representational capacity.

This idea finds a powerful realisation in the framework of \textit{reservoir computing} (RC) \cite{Tanaka2019}. Unlike methods such as DMD - which model dynamics through linear approximations in a predefined basis \cite{williams_datadriven_2015} - RC adopts an agnostic approach. Rather than analytically defining a feature map \( \phi \), RC relies on the intrinsic dynamics of a high-dimensional system to generate the feature space. This is achieved by encoding an input sequence $\vec{X}= [\vec{x}_0,\dots,\vec{x}_t,\dots\vec{x}_T ]$ into a dynamical system termed the \textit{reservoir}. This is typically a fixed, nonlinear dynamical system. Typical examples of this include recurrent neural networks (RNN) \cite{dongReservoirComputingMeets2020}, physical platforms  (e.g., a nonlinear optical cavity) \cite{brunnerRoadmapNeuromorphicPhotonics2025,mcmahonPhysicsOpticalComputing2023,pierangeliPhotonicExtremeLearning2021}, or - in the present case - a quantum many-body system. 

Formally, a reservoir computer maps an input \( \vec{x}_t \) to a reservoir state \( \vec{r}_t \):
\begin{equation}
    \vec{r}_{t+1} = \mathcal{R}(\vec{r}_t, \vec{x}_t),
\end{equation}
where \( \mathcal{R} \) defines the internal dynamics of the reservoir. In general, this recursive relationship between reservoir states is expressive of \textit{memory}, i.e. $\vec{r}_t$ depends not only on $\vec{x}_t$ but also on the sequence of prior inputs. From the reservoir state, we then \textit{readout} some set of observables, which we denote $R(\vec{X})=[\phi(\vec{r}_0),\dots,\phi(\vec{r}_t),\dots\phi(\vec{r}_T)]$. Once again, $\phi$ is some nonlinear transformation that is a product of both the reservoir dynamics $\mathcal{R}$ and the protocol employed for reading out observables from the reservoir. The computation of $\vec{y} $ is then performed by applying a linear map $\vec{W}$ to the readout of the reservoir observables:
\begin{equation}
    \bar{\vec{y}}_t = \vec{W} \phi(\vec{r}_t),
\end{equation}
where $\bar{\vec{y}}_t$ is the computation's approximation to the true output $\vec{y}_t$. A key strength of reservoir computing is its decoupling of dynamics and training: the internal dynamics \( \mathcal{R} \) are fixed and untrained, while the output weights \( \vec{W} \) are learned via a computationally tractable, single-step linear regression (corresponding to a convex optimisation problem). Given a sequence of target outputs \( \vec{Y} = [\vec{y}_0, \dots, \vec{y}_T] \) and corresponding reservoir readouts \( R(\vec{X}) = [\phi(\vec{r}_0), \dots, \phi(\vec{r}_T)] \), the goal is to find the weight matrix that minimises the error $\epsilon$ of the reconstructed target:
\begin{equation}
    \min \epsilon = \underset{\vec{W}}{\arg\min} \sum_{t=0}^{T} \left\| \vec{y}_t - \bar{\vec{y}}_t \right\|^2 = \underset{\vec{W}}{\arg\min} \left\| \vec{Y} - \vec{W} R(\vec{X}) \right\|_F,
\end{equation}
where \( \|\cdot\|_F \) is the Frobenius norm \cite{golub_matrix_2013}. Finding the weights $\vec{W}$ is a regression problem, and to prevent overfitting and improve generalisation, one typically uses Tikhonov (ridge) regression \cite{hastie2009elements}:
\begin{equation}
    \vec{W} = \vec{Y} R(\vec{X})^T\left( R(\vec{X}) R(\vec{X})^T + \lambda \mathbb{I} \right)^{-1}.
\end{equation}
Here \( \lambda \) is a regularisation parameter and \( \mathbb{I} \) is the identity matrix. Unless otherwise specified, we considered a ridge parameter \( \lambda=10^{-4} \) across all our numerical simulations. 

Summarising this process, the reservoir encodes the input sequence into a nonlinear, high-dimensional representation \( \phi(\vec{r}_t) \); the learned weights \( \vec{W} \) then linearly combine these features to approximate the desired output. These weights are calculated via regression, the regularisation of which controls overfitting by penalising large weights. This is an especially important consideration when the reservoir is overparameterised or contains redundancy.

The computational capacity of RC relies on three key properties of the reservoir: nonlinear response, high effective dimensionality, and memory of past inputs. Together, these determine how richly temporal input sequences are transformed and how effectively the output layer can approximate target functions. The RC framework imposes few constraints on how data is encoded or extracted — any physical system with the right structural properties may indeed serve as a reservoir. In the quantum setting, this flexibility is expanded further: inputs may be encoded into quantum states, Hamiltonian parameters, or external controls \cite{masur_optical_2022}; readout may involve projective measurement, expectation values, or continuous monitoring. 

While this broad design space allows for a wide range of implementations, it motivates a central question for this work: what is the \textit{minimal} quantum reservoir architecture — requiring the least overhead in control, measurement, and intrinsic memory — still capable of performing meaningful computation? In many QRC schemes proposed to date, it is assumed that one can repeatedly access the quantum state to perform readout without destroying coherence (see e.g. Ref. \cite{nokkala_gaussian_2021}). This assumption is, however, at odds with the projective and irreversible nature of quantum measurement. While partial or weak measurement protocols may offer some degree of mitigation, any implementation that relies on continuous readout inevitably faces the challenge of state reconstruction, often requiring full or partial tomography — an overhead that scales poorly with system size. 

 This means that while encoding input data into the quantum state of a reservoir allows full access to quantum resources such as entanglement, it also imposes significant demands on implementation — chief among them, high-fidelity state preparation for each input. These requirements closely mirror those of conventional quantum computing, and many of the same obstacles—noise, decoherence, imperfect control—reappear in this setting. Moreover, the relationship between a quantum state and the resulting measurement readout is often non-unique: a challenge made explicit in recent demonstrations of state-to-observable degeneracies and the resulting multiplicity of control fields \cite{mccaul_nonuniqueness_2022, mccaul_optical_2021, magann_sequential_2022, masur_optical_2022}.

In this work, we therefore explore a {\it minimal} alternative architecture, where each datum is encoded not in the quantum state of the reservoir, but in the Hamiltonian parameters that govern its evolution. This \textit{Hamiltonian encoding} ensures that distinct input values correspond to distinct system evolutions and, under fixed initial states and observables, uniquely determine outputs. The one-to-one correspondence arises from the deterministic nature of unitary or dissipative evolution under a specified Hamiltonian \cite{mccaul_driven_2020, mccaul_controlling_2020}. Finally, this approach eliminates the need for complex state preparation: the reservoir is always initialised in a fixed reference state, corresponding, for example, to the ground state of the qubit ensemble, and input data is injected by modulating the Hamiltonian. This substantially reduces experimental overhead while still inducing the necessary nonlinear transformation of inputs.

In this minimal scheme, each data point \( \vec{x}_t \) defines a separate instance of the reservoir’s dynamics via a parameterised Hamiltonian, and the reservoir state is initialised identically each time. That is, the reservoir update rule becomes:
\begin{equation}
    \vec{r}_{t+1} = \mathcal{R}(\vec{x}_t),
\end{equation}
where \( \mathcal{R} \) denotes the evolution under the Hamiltonian associated to input \( \vec{x}_t \), and \( \vec{r}_t \) denotes the resulting reservoir readout. In this minimal setting, the reservoir is strictly memoryless: each input $\vec{x}_t$ defines an independent evolution with no dependence on past inputs. The system is re-initialised after each input, and no coherence is retained between reservoir states. This approach not only lifts the requirements for accurate state preparation but is also naturally compatible with quantum information protocols requiring repeated independent measurements to estimate the expectation values of state operators. 

This reflects a fundamental constraint of quantum measurement: once observables are extracted, the measured subsystem undergoes collapse, erasing any trace of prior inputs. In our architecture, this is not incidental but deliberate — memory is excluded by design. Any temporal dependence must be reconstructed at the level of readout, rather than being embedded in the reservoir's dynamics. More generally, measurement partitions the reservoir: the observed subsystem is projected, while any remaining memory resides in the unmeasured degrees of freedom. Our model makes this partition explicit by designating all measured components as readout and discarding coherence between steps. This means the full reservoir readout matrix can be written simply as:
\begin{equation}
\label{eq:R(X)}
    R(\vec{X}) =[\phi'(\vec{x}_0),\dots, \phi'(\vec{x}_T)]
\end{equation}
where $\phi': \mathbb{R}^{M} \rightarrow \mathbb{R}^{K}$ again maps from input data of dimension $M$ to the reservoir readout of dimension $M$ and (due to absence of memory) is simply:
\begin{equation}
    \phi'(\vec{x}_t)=\phi(\mathcal{R}(\vec{x}_t))
\end{equation}
Although this approach abandons temporal continuity between reservoir states, it allows for a remarkably simple implementation—one that requires no explicit state preparation, no feedback, and no reconstruction after measurement. 

\begin{figure*}[h!tbp]
    \centering
    \includegraphics[width=0.9\linewidth]{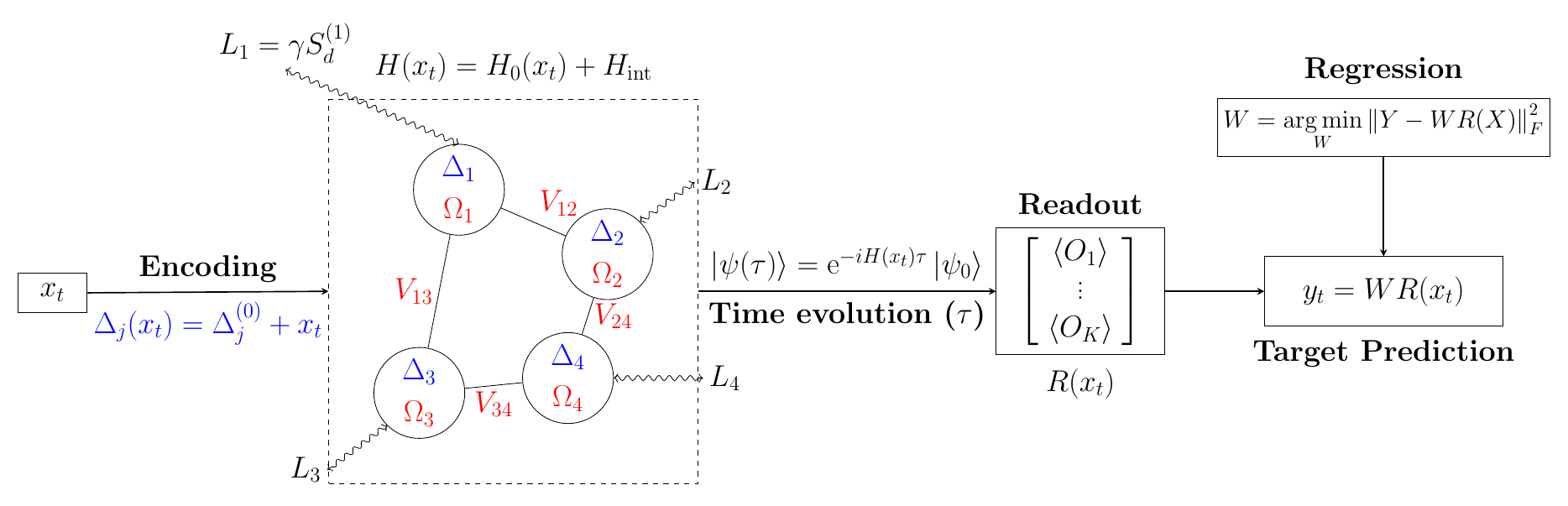}
 \caption{
    Schematic representation of the quantum reservoir computing (QRC) architecture explored in this work. 
    A one-dimensional input stream \( x_t \in \mathbb{R} \) is encoded into the dynamics of a driven-dissipative quantum system by modulating local detuning terms, \( \Delta_j(x_t) = \Delta_j^{(0)} + x_t \), in a Hamiltonian of the form \( H(x_t) = H_0(x_t) + H_{\mathrm{int}} \). The system evolves according to the dynamics outlined in the main text, and at a fixed final time \( \tau \), a set of observable expectations \( \langle O_k \rangle \) is measured to yield the reservoir state vector \( R(x_t) \). This state is then passed to a linear readout layer, which is trained via ridge regression to predict target outputs \( y_t \) using \( y_t = W R(x_t) \).
    }
    \label{fig:qrc_schematic}
\end{figure*}
% {\color{red} To illustrate our approach, we consider a $N$-qubit reservoir governed by a generic Ising-type Hamiltonian, which represents a wide class of quantum systems ranged form cold atoms to superconducting qubits.}
\subsection*{Model}
To explore the computational capacity of Hamiltonian encoding, we consider a minimal quantum reservoir composed of \( N \) qubits evolving under a driven-dissipative Hamiltonian. The system is inspired by Ising-type models \cite{andreevEmergenceControlComplex2021}, and captures a broad class of experimentally relevant platforms ranging from cold atoms to superconducting qubits \cite{mohseni_ising_2022}.

Each input \( x_t \in \mathbb{R} \) is mapped onto a time-independent Hamiltonian \( H(x_t) \), which governs the evolution of the reservoir over a fixed interval \( \tau \). We decompose the Hamiltonian into a local driving term and a symmetric two-body interaction:
\begin{equation}
    \label{eq:ham0}
    H(x_t) = H_0(x_t) + H_{\mathrm{int}},
\end{equation}
where
\begin{equation}
    \label{eq:ham1}
    H_0(x_t) = \sum_{j=1}^{N} \left[-\Delta_j(x_t)\, S_d^{(j)} + \frac{\Omega_j}{2}\, \sigma_x^{(j)} \right],
\end{equation}
and
\begin{equation}
    \label{eq:ham2}
    H_{\mathrm{int}} = \sum_{m<n} \frac{V_{m,n}}{N-1}\, S_d^{(m)} S_d^{(n)}.
\end{equation}
Here, the operator \( S_d^{(j)} = \tfrac{1}{2}(I - \sigma_z^{(j)}) \) projects onto the excited state of qubit \( j \), and the scaling of the interaction ensures energy extensivity. The detuning terms are modulated by the input signal as
\begin{equation}
    \Delta_j(x_t) = \Delta_j^{(0)} + s x_t,
    \label{eq:scal}
\end{equation}
where \( \Delta_j^{(0)} \) denotes a base detuning parameter, while $s$ is a scaling hyperparameter.
%that is randomly drawn for each qubit to introduce heterogeneity into the reservoir, while $s$ is a scaling hyperparameter.
 To introduce coupling heterogeneity into the reservoir, both  \( \Delta_j^{(0)} \) and the Rabi frequencies \( \Omega_j \) are randomly sampled from a normal distribution around the average $\Delta_0$ and $\Omega_0$, with the width of this distribution corresponding to $\sim 10 \%$ of these average values.

%by drawing from uniform distributions centred on $\Delta_0$ and $\Omega_0$. Similarly the (symmetric) interaction amplitudes are also \( V_{m,n} \) chosen randomly from a uniform distribution around $V_0$.  

Dissipation is incorporated via Lindblad operators corresponding to spontaneous emission on each qubit, i.e. $\hat{L}_j = \gamma\hat{S}^{(j)}_d$ and for each input the system is evolved over a fixed time window $\tau$ from a common (but randomly selected) initial state $\ket{\psi_0}$. To model a more realistic experimental scenario, the reservoir readout is then obtained by taking observable expectations at this final time. 
 
Explicitly, we have:
\begin{equation}
    \phi(\vec{r}_t)= [O_1(x_t),\dots O_K(x_t)]^T,
\end{equation}
where 
\begin{equation}
    O_j(x_t)= \bra{\psi_0}{{\rm e}}^{i\hat{H}(x_t)\tau}\hat{O}_j{{\rm e}}^{-i\hat{H}(x_t)\tau}\ket{\psi_0}
\end{equation}
and the operators $\{\hat{O}_j\}$ correspond to some set of system observables, e.g. \( \{\sigma_z^{(j)}\} \).  The resulting reservoir readout \( R(x_t) \) is constructed from these expectation values. The full $R(X)$ is constructed by independently repeating this process for
each element of the input time series, and is then supplied to a linear readout layer for regression. 

As a practical matter, this specification of independent processing will greatly reduce fidelity and coherence time demands on hardware, as compared to sequential processing in a single evolution. The cost of this is that the response generated by the reservoir will be \textit{memoryless}, i.e. the expectations corresponding to each datum have no dependence on the sequence of inputs preceding it. This setup is schematically represented in Fig. \ref{fig:qrc_schematic}, illustrating the entire pipeline from encoding to model prediction.

\subsection*{Distinctions between state and Hamiltonian encoding}
Within this memoryless set-up, it is possible to illustrate the distinction between \textit{state} and \textit{Hamiltonian} encoding. Let us assume the system Hamiltonian depends linearly on the scalar input \( x_t \) via
\begin{equation}
    \hat{H}(x_t) = \hat{H}_0 + x_t \hat{H}_x.
\end{equation}
This class of models includes the particular Hamiltonian outlined in Eqs. \eqref{eq:ham0}-\eqref{eq:ham2} and incorporates any setup in which data can be encoded via the amplitude of some control terms. 

For illustrative purposes, we take \( \hat{H}_0 \) to be non-degenerate with eigenvalues \( E^0_j\) and eigenstates $\ket{j}$, satisfying \( \hat{H}_0 \ket{j_0} = E^0_j \ket{j_0}\). We further label the corresponding $\hat{H}(x_t)$ eigenvalues $E_j^{x_t}$ with eigenstates \( |j_{x_t} \rangle \). With this established, we can consider two scenarios for generating the final state $\ket{\psi_\tau(x_t)}$ from which the reservoir readouts are obtained. That is, using either the previously outlined Hamiltonian encoding:
 \begin{equation}
\ket{\psi_\tau(x_t)} = {\rm e}^{-i\hat{H}(x_t)\tau}\ket{\psi_0},
 \end{equation}
 or an \textit{initial state} encoding $\ket{\psi_0(x_t)}$ evolving under a fixed $\hat{H}_0$:
 \begin{equation}
     \ket{\psi_\tau(x_t)} = {\rm e}^{-i\hat{H}_0\tau}\ket{\psi_0(x_t)},
 \end{equation}
 we can infer the state encoding necessary to render an equivalent readout state to the Hamiltonian encoded system. Defining our state encoding in the basis of $\hat{H}_0$ as:
 \begin{equation}
  \ket{\psi_0(x_t)} = \sum_j c_j(x_t) \ket{j_0}   
 \end{equation}
 such that 
 \begin{equation}
     \ket{\psi_\tau(x_t)} = \sum_j {\rm e}^{-i\tau E^0_j} c_j(x_t) \ket{j_0}.  
 \end{equation}
 In the Hamiltonian encoding scenario, we will similarly express our random state in the basis of $\hat{H}_0$ (so that its $c_j^0$ amplitudes are $x_t$ independent:
  \begin{equation}
  \ket{\psi_0} = \sum_{j} c^0_j  \ket{j_0}.  
 \end{equation}
Expressing the final state in the unperturbed basis then yields:
  \begin{equation}
  \ket{\psi_\tau (x_t)} = \sum_{jkl} c^0_j  {\rm e}^{-i\tau E^{x_t}_k} \braket{k_{x_t}|j_0}\braket{l_0|k_{x_t}} \ket{j_0}.  
 \end{equation}
 For the two methods to be equivalent, the state encoding must therefore follow: 
 \begin{equation}
     c_j(x_t)=\sum_{kl} c^0_j  {\rm e}^{-i\tau (E^{x_t}_k-E^0_j)} \braket{k_{x_t}|j_0}\braket{l_0|k_{x_t}}.
 \end{equation}

This expression illustrates the structural complexity required of a state encoding to express an equivalent Hamiltonian encoding. Even to leading order, the coefficient $c_j(x_t)$ is a nonlinear function of $x_t$, mediated by the eigenvalues $E_k^{x_t}$ and overlaps involving $\ket{k_{x_t}}$, both of which depend implicitly on $x_t$. Expanding each term perturbatively, we find
\begin{align}
    E_k^{x_t} &= E_k^{0} + x_t \bra{k_0} \hat{H}_x \ket{k_0} + \mathcal{O}(x_t^2), \\
    \braket{k_{x_t}|j_0} &= \delta_{kj} + x_t \frac{ \bra{j_0} \hat{H}_x \ket{k_0} }{ E_k^{(0)} - E_j^{(0)} } + \mathcal{O}(x_t^2),
\end{align}
with a similar expansion for $\braket{l_0|k_{x_t}}$. Substituting into the expression for $c_j(x_t)$ reveals nested rational functions and phase modulations with $x_t$-dependent frequencies, along with interference terms that depend on spectral gaps and operator overlaps. As a result, $c_j(x_t)$ exhibits an oscillatory, non-polynomial structure in both $x_t$ and the (hyperparameter) evolution time $\tau$.  

Moreover, the structure of $c_j(x_t)$ will \textit{also} depend on both the relative size and eigenbases of $\hat{H}_0$ and $\hat{H}_x$. The former property will control how many relevant orders of perturbation (and therefore polynomial degrees of $x_t$) contribute, while the commutation structure of the operators will determine mixing terms. Specifically, when $[\hat{H}_0, \hat{H}_x] = 0$, the eigenbases coincide and (to first order in $x_t$) $c_j(x_t) = c_j^0 \, {\rm e}^{-i\tau x_t \bra{j_0} \hat{H}_x \ket{j_0}}$, corresponding to a simple phase encoding. However, when $[\hat{H}_0, \hat{H}_x] \ne 0$, off-diagonal mixing between eigenstates introduces amplitude reshaping and multi-frequency phase interference, which compounds with increasing $\tau$. The effective encoding map thus becomes strongly nonlinear, even in the absence of memory or feedback.

%Concluding this discussion, it is worth pausing to consider the trade-off between encoding dimensionality and non-linearity inherent in state versus Hamiltonian encoding. In an $n$-site, $k$-local Hamiltonian, Hamiltonian encoding permits control over only $\mathcal{O}(n^k)$ parameters. By contrast, state encoding allows the preparation of arbitrary vectors in a $2^n$-dimensional Hilbert space, enabling an exponentially greater input dimensionality—but under the constraint of encoding schemes which are experimentally preparable and distinguishable. In short, Hamiltonian encoding guarantees a nonlinear transformation of input data in a compact manner, while state-encoding allows for the embedding of much higher-dimensional data where amplitude non-linearity must be encoded by hand. 

The above analysis suggests that when considering the trade-off between encoding dimensionality and non-linearity in state versus Hamiltonian encoding, the former allows control over only $\mathcal{O}(n^k)$ parameters in an $n$-site, $k$-local Hamiltonian. Conversely, state encoding enables the preparation of arbitrary rays in a $2^n$-dimensional Hilbert space. This allows for exponentially greater input dimensionality, but can only be exploited under the constraint of experimentally preparable and distinguishable encoding schemes. 

In summary, Hamiltonian encoding guarantees a compact nonlinear transformation of input data, while state encoding allows for the embedding of much higher-dimensional data where amplitude non-linearity must be encoded manually.

%  Then, to leading orders in perturbation theory, the eigenvalues \( E_j^{(x)} \) and eigenstates \( |j^{(x)}\rangle \) of the perturbed Hamiltonian \( \hat{H}(x_t) \) are given by:
% \begin{align}
%     E_j^{(x)} &= E_j^{(0)} + x_t \langle j^{(0)} | \hat{H}_x | j^{(0)} \rangle + x_t^2 \sum_{k \neq j} \frac{|\langle k^{(0)} | \hat{H}_x | j^{(0)} \rangle|^2}{E_j^{(0)} - E_k^{(0)}} + \mathcal{O}(x_t^3), \label{eq:eigenvalue_pert}\\
%     |j^{(x)}\rangle &= |j^{(0)}\rangle + x_t \sum_{k \neq j} \frac{\langle k^{(0)} | \hat{H}_x | j^{(0)} \rangle}{E_j^{(0)} - E_k^{(0)}} |k^{(0)}\rangle + \mathcal{O}(x_t^2). \label{eq:eigenstate_pert}
% \end{align}
% If the initial state 

\subsection*{Augmenting Reservoir Capacity with Delay Embeddings}
\begin{figure}[h!tbp]
    \centering
    \includegraphics[width=0.8\linewidth]{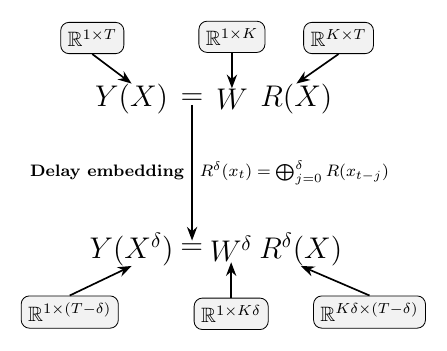}
 \caption{
Schematic illustration of delay embedding applied at the readout level of a quantum reservoir computer. The top row shows the standard formulation in which a reservoir readout matrix \( R(X) \in \mathbb{R}^{K \times T} \) is mapped to the output \( Y(X) \in \mathbb{R}^{1 \times T} \) via a learned weight matrix \( W \in \mathbb{R}^{1 \times K} \). The bottom row shows the result of applying post-processing delay embedding to the readout. This extends each \( R(x_t) \) by concatenating it with \(\delta\) delayed versions of itself, producing an augmented readout matrix \( R^\delta(X) \in \mathbb{R}^{K\delta \times (T - \delta)} \). The associated regression matrix \( W^\delta \in \mathbb{R}^{1 \times K\delta} \) now operates on this extended feature space to produce a prediction over the shortened time axis \( T - \delta \). This construction enables predictive performance even for reservoirs lacking intrinsic memory.
 }
    \label{fig:delay_reshape}
\end{figure}
As outlined previously, this model foregoes any explicit memory in the reservoir. Implicitly, this greatly limits its expressive capacity, as one of the paradigmatic tasks in reservoir computing (and machine learning generally) - \textit{forecasting or prediction} - relies on memory. If our input data is given by a time series $u(t)$ such that $x_t=u(t)$, the prediction task is defined by a target $y_t = u({t+\kappa})$, i.e. given the time-series at $t$, predict its value at time $t+\kappa$. In general, this is a qualitatively different task from synthesis, as there is no guarantee of a bijective correspondence between $x_t=x(t)$ and $x_{t+\kappa}=x(t+\kappa)$. 

It is for this reason that a reservoir requires memory to make predictions. In the present case, the minimal encoding and readout scheme of our quantum reservoir has been designed to explicitly \textit{exclude} memory effects. This would \textit{prima facie} imply that the minimal reservoir proposed here lacks the expressivity necessary to perform prediction. There is, however, another perspective on this problem, motivated by Takens' \textit{embedding theorem} \cite{Takens,timeseriesgeometry}. This states that any dynamical system whose observables (say for example $x(t) \in \mathbb{R}$) evolve smoothly, then for any fixed delay \(\tau > 0\), there exists a delay-coordinate map 
\[
\Phi(x(t)) = \Bigl(x(t), x(t-\kappa), x(t-2\kappa)..., x(t-m\kappa) \Bigr)
\]
that defines an \textit{embedding} of the observable dynamics into $\mathbb{R}^{m+1}$.

% This observation opens a compelling route for enhancing predictive capacity without modifying the reservoir itself. Rather than introducing internal feedback or memory into the reservoir dynamics—a step which would complicate both the physical implementation and the interpretability of the system—we instead propose to augment the reservoir's output layer using a \textit{delay embeeding} scheme. [REFS HERE].  Concretely, we extend the reservoir readout vector \( R(x_t) \to R^\delta(x_t)\) by appending delayed versions of the original input signal. That is, we define an augmented readout:
% \begin{equation}
%     R^\delta(x_t) = \bigoplus_{j=0}^{j=\delta}R(x_{t-j})
% \end{equation}
% where $\delta$ controls the number of delayed copies. This construction mirrors the delay-coordinate embedding described by Takens' theorem, but applies it directly at the level of readout rather than state. Importantly, the reservoir evolution remains memoryless and input-local — each \( x_t \) is still processed independently via Hamiltonian dynamics — but allows the temporal dependencies \textit{of the input itself} to be encoded into the augmented reservoir dimension. This strategy of \emph{post-processing delay embedding}, allows even a stateless reservoir to perform nontrivial prediction tasks without sacrificing the minimality or modularity of the quantum implementation.

This observation suggests a simple route to enhancing the predictive capacity of a reservoir \textit{without} requiring it to exhibit intrinsic memory. Rather than introducing internal feedback or memory into the reservoir dynamics—a step that would complicate both the physical implementation and theoretical clarity—we instead propose a \textit{delay embedding} strategy applied entirely at the level of readout. This elaboration has been used extensively to augment reservoir expressivity in a wide variety of contexts \cite{Lu2017,Gauthier:2021aa,Jaurigue:2025aa}, but this has usually been used to enhance memory properties already present in the reservoir. In the present case, however, delay embedding endows the reservoir with a property that would otherwise be entirely absent. 

Concretely, we extend the reservoir readout vector \( R(x_t) \) to an augmented form \( R^\delta(x_t) \) by appending delayed versions of the input:
\begin{equation}
    R^\delta(x_t) = \bigoplus_{j=0}^{\delta} R(x_{t-j}),
\end{equation}
where \( \delta \) controls the number of delays included and \( \bigoplus \) denotes concatenation by direct sum. The corresponding reshaping is outlined in Fig. \ref{fig:delay_reshape}.  This construction mirrors the delay-coordinate embedding described by Takens' theorem, but shifts its application from state to output. Crucially, the reservoir dynamics remain memoryless and input-local—each \( x_t \) is processed independently under its associated Hamiltonian—but temporal correlations in the input sequence are reintroduced at the level of regression. This strategy, termed \emph{post-processing delay embedding}\cite{Jaurigue:2025aa}, enables even a memoryless reservoir to perform nontrivial prediction tasks without compromising the minimality or interpretability of the underlying quantum architecture.

\section{Results and Discussion \label{sec:results}}
Using the minimal reservoir outlined, we now examine its performance across various tasks, both with and without delay augmentation. Unless otherwise specified, in all results we use a reservoir of \( N = 5 \) qubits, as this represents the minimal system size that consistently achieves low regression error across nonlinear targets. The base detuning is set to \( \Delta_0 = 5 \), with Rabi frequencies centered at \( \Omega_0 = 2 \). Each input \( x_t \) defines a distinct Hamiltonian evolution over a fixed time interval \( \tau = 1.5\pi / \Omega_0 \). Dissipation is included via Lindblad operators with rate \( \gamma = 1.5 \times 10^{-2} \). The datasets employed span \( N_T = 2500 \) inputs, ensuring stable convergence of the regression across all tasks. These parameters generate sufficiently heterogeneous dynamics to support nonlinear regression while remaining within the tractable regime for simulation and potential implementation.

To characterise model performance, we use the Normalised Mean Square Error (NMSE), defined as \cite{Jaeger2004}
\begin{equation}
    \mathrm{NMSE} = \frac{\sum_{t} \left( y_t - \bar{y}_t \right)^2}{\sigma^2(y)}
\end{equation}
where \( y_t \) is the target, \( \bar{y}_t \) the predicted output, and $\sigma^2(y)$ the variance of the target sequence. Lower NMSE values indicate more accurate regression, with \( \mathrm{NMSE} = 0 \) corresponding to perfect prediction.

\subsection{Nonlinear Regression}

\begin{figure}[h!tbp]
    \centering
     \includegraphics[width=0.7\linewidth]{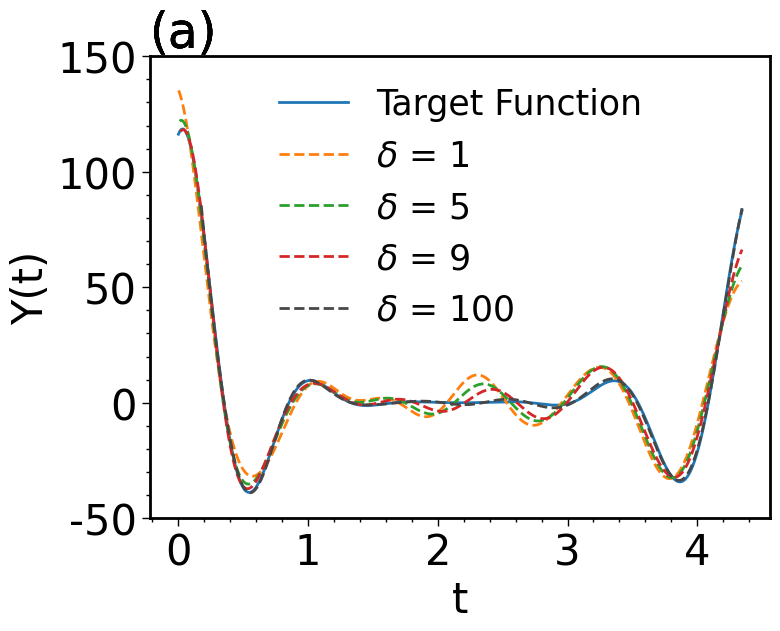} \\
     \includegraphics[width=0.7\linewidth]{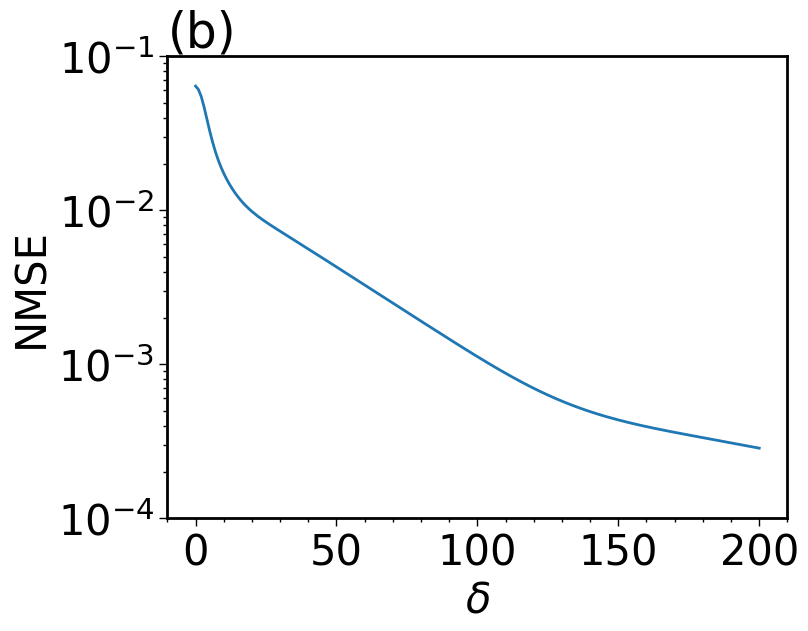}
 \caption{Example regression for a trigonometric-polynomial target function $y_t=f(t)$, where $f(t)=t \sin(t) + t^2 \cos(2t-0.2) + t^3 \sin(5t+0.4) + t^4 \cos(4t-0.3)$. Here we use $N = 5$ qubits and scale $s = 0.8$, and find that the Normalised Mean Square Error (NMSE) is improved montonically as the number of incorporated delays $\delta$ is increased.  
    \label{fig:delays_example}}
\end{figure}

 As a first task, we take an input $x_t$ over a regular domain, and consider regression for typical polynomial and trigonometric functions, i.e. $y_t =f(x_t)$. Importantly, the ability of the reservoir to perform such tasks is dependent on both the physical and encoding parameters, and serves as effective hyperparameters for the regression model. Importantly, while these parameters can be tuned to a given task, in the absence of delays, the minimal reservoir is not generically performant in synthesising target functions. By including delay embeddings, we find that reservoir expressivity is both increased and its sensitivity to the choice of hyperparameters is greatly reduced. 

 This finding is illustrated in Fig. \ref{fig:delays_example}, where a trigonometric polynomial function is used as a synthesis target for the reservoir with $N=5$. Figure  \ref{fig:delays_example}(a) demonstrates how the prediction of the target function (solid line) evolves with variation of the delays $\delta$  in post-processing embedding. As $\delta$ increases, the prediction more closely aligns with the target. This improvement is confirmed by Fig. \ref{fig:delays_example}(b), where NMSE diminishes exponentially with $\delta$.

\begin{figure}[h!tbp]
    \centering
     \includegraphics[width=8.5cm]{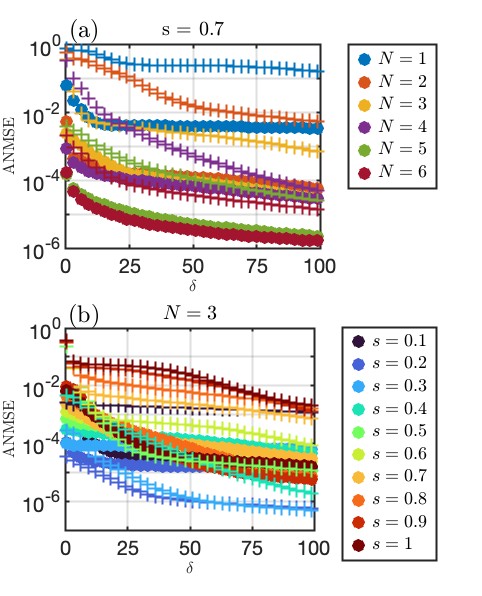}
 \caption{\label{fig:delayANMSE} Average NMSE (ANMSE) decreases with increasing number of delays for a) constant scale and varied $N$ (scale = 0.7) and b) constant $N$ and changing scale ($N = 3$). The ANMSE of two groups are considered: Group 1 - $u$, $u^2$ and $u^3$ shown by circle markers and Group 2 - $sin(u)$ and $cos(u)$ marked by $+$. 
  }
\end{figure}

The impact of reservoir size \( N \) and input scaling \( s \) [see Eq.~(\ref{eq:scal})] on regression accuracy is summarised in Fig.~\ref{fig:delayANMSE}, which shows how performance varies with the number of delay embeddings \( \delta \). Two groups of target functions $y_t = u(t)$ are considered: polynomial functions \( u(t) = t,\, t^2,\, t^3 \), and trigonometric functions \( u(t) = \sin(t),\, \cos(t) \). For each configuration, we compute the Normalised Mean Square Error (NMSE) between predicted and target outputs, and report the average NMSE (ANMSE) across each function group. Panel (a) displays the ANMSE as a function of \( \delta \) for various values of \( N \) at fixed input scaling \( s \), while panel (b) shows the same quantity for varying \( s \) at fixed \( N \). 

The results show that for \( N \geq 3 \), an ANMSE below \( 10^{-3} \) can be achieved for both function groups, provided the number of delays is sufficiently large (typically \( \delta \gtrsim 50 \)). Polynomial targets can be approximated accurately even without delay embedding when \( N \geq 3 \), while trigonometric functions require a significantly larger \( \delta \) to reach comparable accuracy. This suggests that alternating (non-monotonic) functions require a greater effective dimensionality of the reservoir to accurately synthesise.

Figure~\ref{fig:delayANMSE}(b) also illustrates how the scaling of the input values $s$ impacts ANMSE for a range of $\delta$. In particular, as $\delta$ increases, ANMSE significantly decreases for all values of $s$. Thus, expanding dimensionality via post-processing embedding substantially reduces the fitting’s sensitivity to the input scaling.

% %This improvement with delay embeddings is illustrated more generally in Fig. \ref{fig:delayANMSE}, where we find 
%The graphs clearly indicate that the average fitting performance over both polynomial and trigonometric test functions is uniformly improved. The precise degree of improvement is however dependent on the underlying dimensionality of the system. In particular, we observe in Fig. \ref{fig:delayANMSE} \textbf{a)} that for $N\geq 3$ the reservoir is able to accurately reconstruct polynomial functions without delays, but requires $N\geq 5$ to achieve similar accuracy for fitting to trigonometric targets. An underlying sensitivity of the system to the scaling of input $s$ in (\ref{eq:scal}) is also observed in Fig. \ref{fig:delayANMSE} \textbf{b)}, but once again the uniform improvement of performance with respect to the number of delays embedded diminishes the need for such fine tuning of hyperparameters. Finally, we see direct evidence for the additional expressivity conferred by larger system sizes - across test functions we observe that for larger $N$, a smaller $\delta$ is sufficient to achieve a given fidelity (for example {\color{red} averaged NMSE (ANMSE)} $\sim 10^{-3}$) to the target.  {\color{red} {\bf !!! We need more detailed discussion of Fig.4, especially lower panels (a)-(d)!!!}}

\subsection{Time-dependent tasks: synthesis and prediction with memory-less QRC systems}

\begin{figure}[h!]
    \includegraphics[width=\linewidth]{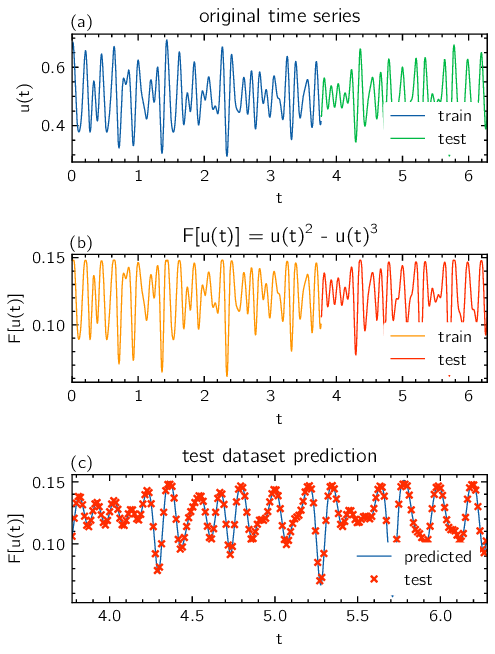}
    \caption{\label{fig:timeseries} (a): Example of time series generated according to Eq.\eqref{eq:randomsin} split into training (blue line) and test (green line) datasets. (b): The target function $F(u)=u^2-u^3$ used as for the synthesis tasks. (c): Prediction results (blue line) vs ground truth values (red cross) for the dataset in panel b. Simulation parameters: $N_{qubit}=5, \gamma = 10^{-8}, \Delta_0 = 5, \Omega_0 = 2$.}
\end{figure}

Having considered this minimal reservoir's capacity for regression, we now investigate its ability to perform time series synthesis and prediction. Time-series analysis is a cornerstone of reservoir computing methods, particularly when considering prediction/forecasting tasks. Intuitively, one would expect that the lack of intrinsic memory in our QR protocol, originating from resetting the initial state for each data point, would compromise the ability to perform time-dependent tasks. However, the introduction of post-processing delay embedding enables engineering an artificial memory effect by expanding the number of dimensions at the readout of the reservoir system, effectively emulating the effect of intrinsic memory in time-dependent reservoirs. Unlike standard approaches, the combination of Hamiltonian encoding and delay embedding enables us to perform time-dependent tasks, even under independence of data processing - a feature highly desirable for hardware implementations. 

To demonstrate this, we consider a synthetic input signal defined as a sum of sinusoids with random, non-commensurate frequencies and phases:
\begin{equation}
    \label{eq:randomsin}
    u(t) = \sum_{j=1}^5 \sin(\omega_j t - \phi_j),
\end{equation}
where \( t \in [0, 2\pi] \). The frequencies \( \omega_j \) are sampled from a Gaussian distribution \( \mathcal{N}(2 f_0, f_0) \) with \( f_0 = 20 \), and the phases \( \phi_j \) are drawn uniformly from \( [0, 2\pi] \). This construction combines deterministic (sinusoidal functions) and random structure, as the sum of incommensurate sinusoids yields a quasi-random signal. 

We discretise this signal over \( N_T = 1000 \) uniformly spaced time points with step size $\Delta$, and denoting the sampled sequence by \( u_j = u(t=j \Delta) \). The dataset is split into training and test subsets of size \( N_{\text{train}} = 600 \) and \( N_{\text{test}} = 400 \), respectively.

As an initial task, we consider a pointwise nonlinear transformation of the input sequence, defined by the target $y_j=Y(t=j\Delta)$:
\begin{equation}
    y_j = F(u(j\Delta)) = u_j^2 - u_j^3.
\end{equation}
This synthesis task mirrors the regression setting discussed previously, applied to temporally ordered data. Example results of this synthesis are shown in Fig. \ref{fig:timeseries}, with the realised time-series shown in panel \ref{fig:timeseries}(a), and the corresponding nonlinear target $F[u(t)]$ in \ref{fig:timeseries}(b). Fig. \ref{fig:timeseries}(c) demonstrates a QR with $N=5$ qubits is able to perform the task with a remarkable degree of accuracy, yielding training and test NMSE values of 3.59$\times$10$^{-3}$ and 1.10$\times$10$^{-3}$ respectively. As expected, in this time-local task, the native reservoir performs well, successfully predicting a range of synthesized signals even without introducing any additional delay embedding. 

To assess the predictive capabilities of our model, we then considered time-series prediction tasks. As in standard practice, time-series prediction tasks can be divided into non-autonomous and autonomous prediction. In both scenarios, the reservoir is trained to predict the value of the input function at a future time step, namely   
\begin{equation}
    \label{eq:tspred}
    u(t) \longrightarrow Y(t) = u(t+\kappa), 
\end{equation}
where $\kappa =k\Delta$ is the time offset, expressed in terms of the number of discrete steps $k$ into the future. While the training phase is analogous for both cases, in the non-autonomous case (for prediction offset $\kappa$), the reservoir is fed as an input the current value of the time-series: 
\begin{equation}
    \label{eq:nonautopred}
    \bar{y}_j=WR(y_{j-k}).
\end{equation}
In the autonomous case, the reservoir output at time $t=j\Delta$ is used as a feedback input to the next time-step $t=(j+1)\Delta$, i.e.
\begin{equation}
    \label{eq:autopred}
    \bar{y}_j=WR(\bar{y}_{j-1}).
\end{equation}

\begin{figure*}[h!btp]
    \centering
    \includegraphics[width=\linewidth]{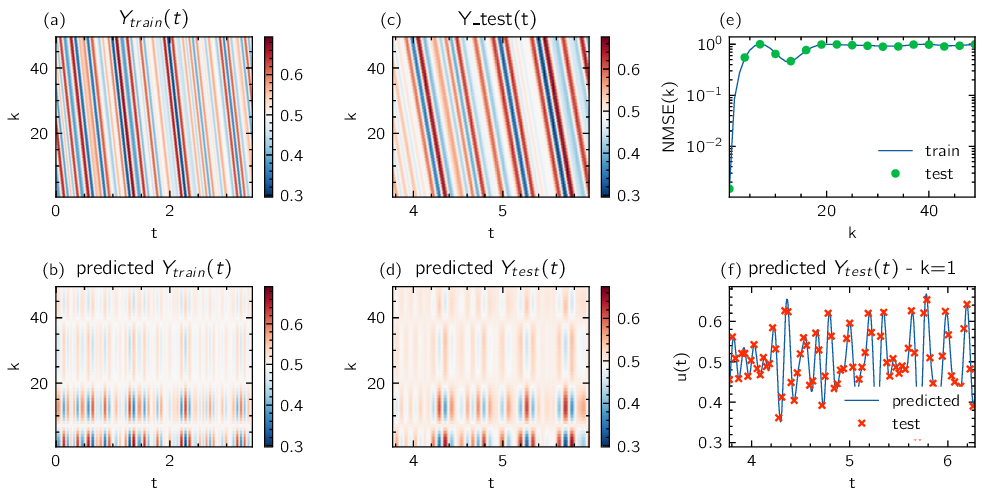}
    \caption{\label{fig:non-autonomous} \textbf{Non-autonomous time-series prediction.} (a,b) Ground truth (panel a) and predicted (panel b) time-series $Y(t_n)$ for different values of the prediction offset $\kappa = k \Delta$ in the non-autonomous prediction task. The pseudo-color plots correspond to the training dataset. (c,d) Same as panels a and b, but considering the test dataset. (e) Normalized Mean-Square Error (NMSE) for the training (solid blue line) and test (green dots) dataset as a function of the prediction offset $\kappa$. (f) Example of predicted timeseries for $k=1$. Simulation parameters: $N_{qubit}=5, \gamma = 10^{-8}, \Delta_0 = 5, \Omega_0 = 2$.}
\end{figure*}

Fig. \ref{fig:non-autonomous} reports results for the non-autonomous prediction for different values of the prediction offset, illustrating the limitations of the native reservoir due to the lack of an intrinsic memory mechanism. Panels \ref{fig:non-autonomous}(a,b) show the training ground-truth and the corresponding predicted values of the output target $Y(t)$, respectively. Figure \ref{fig:non-autonomous}(c,d) corresponds to the same quantities for the test dataset. In all four panels, the pseudo-colour map corresponds to the target (a,c) and predicted (b,d) time-series as a function of the prediction offset $k$ and the discretised time $t$. Interestingly, the performance of the reservoir system does not particularly vary across the two datasets, as illustrated in Fig. \ref{fig:non-autonomous}(e), where we report the two NMSE values (solid blue line and green dots) as a function of the number of discrete steps $k$ in the prediction offset $\kappa=k\Delta$. In both cases, the reservoir appears to predict a non-shifted version of the input time-series up to some critical $\kappa$, after which the predicted output $\bar{y}_t$ undergoes periodic phase shifts with respect to $\kappa$ [see Fig. \ref{fig:non-autonomous}(b,d)]. For small values of the prediction offset $\kappa$, the output time series is in fact almost indistinguishable from the input time series. These results emphasise the critical role of memory in prediction, underscoring the fundamental limitation imposed by its absence. 

\begin{figure*}[h!tbp]
    \centering
    \includegraphics[width=\linewidth]{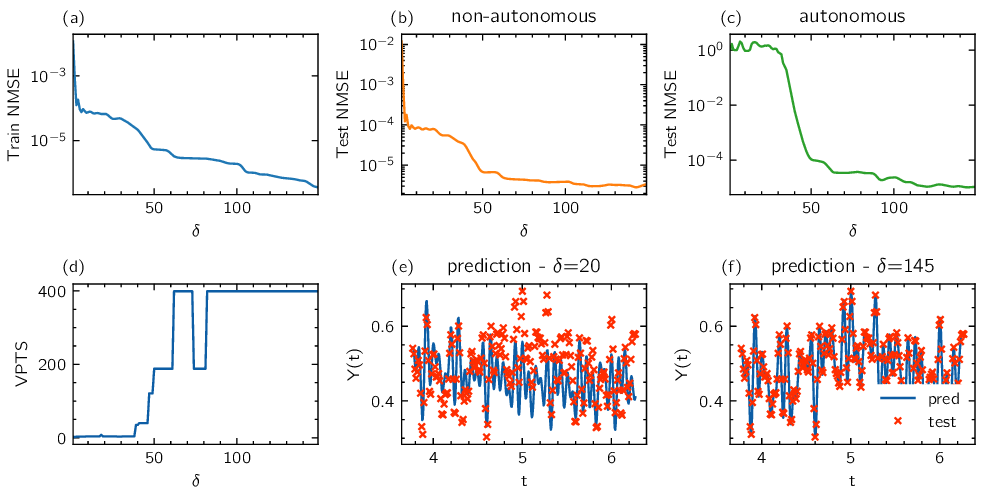}
    \caption{\label{fig:delay-performance} \textbf{Time-series prediction via delay embedding.} (a-c) Normalized Mean Square Error (NMSE) for the training (panel a), non-autonomous test (panel b), and autonomous test (panel c) as a function of the number of delay embeddings $\delta$. The input time series considered here is the same as in Fig. \ref{fig:timeseries}(a).  In all cases, the NMSE decreases monotonically as the number of delays increases, reflecting the improved predictive capacity conferred by delay augmentation. (d) Valid Prediction Time Steps (VPTS) as a function of the number of delay embeddings $\delta$. (e,f) Example of autonomous predictions on the test dataset using $\delta=20$ (panel e) and $\delta=145$ (panel f) delays, respectively. Simulation parameters: $N_{qubit}=5, \gamma = 10^{-8}, \Delta_0 = 5, \Omega_0 = 2$.
    }
\end{figure*}

We further investigate the predictive capabilities of our minimal QR system by introducing the delay-embedding technique discussed in the previous section. The results are shown in Fig. \ref{fig:delay-performance}. To quantify how performance depends on the number of delay embeddings $\delta$, we report in Fig.~\ref{fig:delay-performance}(a)–(c) the model’s single-step prediction error for the training, non-autonomous, and autonomous test stages, respectively. We stress here that the input time series was processed only once through the QR, and all the delay embedding operations were performed by considering the same QR output in post-processing. Our results exhibit a consistent pattern across all cases, with the NMSE decreasing monotonically as the number of delays increases; the most dramatic improvement occurs beyond approximately 30–50 delays. This trend reflects the role of delay embeddings in effectively reconstructing the underlying state space, even when the reservoir itself lacks intrinsic memory. 

To quantitatively estimate the autonomous prediction performance, we computed the valid prediction time steps (VPTS) metric, as shown in Fig. \ref{fig:delay-performance}(d). The VPTS is defined as the number of steps at which the autonomous NMSE testing error exceeds the value 0.4, and it quantifies the reservoir's ability to remain on the true trajectory of the input time series
\cite{kosterAttentionenhancedReservoirComputing2024,kosterDatainformedReservoirComputing2023}:
\begin{equation}
    \label{eq:vpts}
    t_V = \max_t{ | NMSE_{{\rm test}}(t)| < 0.4}
\end{equation}
In practice, the VPTS measures the number of consecutive time steps over which the autonomous prediction remains stable and accurately tracks the target trajectory before diverging or collapsing to a trivial solution. This provides a practical measure of long-term forecast stability, complementing global error metrics such as the NMSE. As the number of delays increases, the VPTS exhibits sharp transitions, indicating critical thresholds beyond which the system achieves long-term stability. This is consistent with the idea that delay embeddings extend the effective dimensionality of the reservoir, enabling it to reconstruct attractor geometry and support stable autonomous forecasts \cite{timeseriesgeometry, Takens}. 

Taken together, the regression, synthesis, and prediction results highlight a unifying principle: while the minimal Hamiltonian-encoded reservoir lacks intrinsic memory, it nevertheless produces a high-dimensional, nonlinear embedding of the input that enables meaningful regression and synthesis tasks. However, for temporal prediction, the absence of memory imposes a fundamental constraint. By augmenting the readout with delay embeddings, the system receives the necessary temporal target information directly. This can be interpreted as lifting the reservoir into a higher-dimensional space where the attractor geometry underlying the target series can be reconstructed.  The combined results thus demonstrate that predictive capacity in a memoryless reservoir is fundamentally a matter of dimensionality, and that delay embeddings provide a principled and efficient route to overcoming architectural minimalism.

\section{Conclusions \label{sec:discussion}}

This work introduces and characterizes a minimal quantum reservoir computing architecture that encodes input data directly into the Hamiltonian parameters of a dissipative quantum system. Our architecture is minimal in the sense that it removes all features not essential to the quantum system's operation as a reservoir. This deliberately eschews memory, recurrent evolution, feedback, and explicit state entangling, retaining only a fixed initial state, a parameter-dependent Hamiltonian, and a projective readout after evolution. By eliminating the need for state preparation, feedback, or tomography, this approach offers an experimentally lightweight and conceptually transparent platform for quantum machine learning. 

Despite its lack of intrinsic memory, we have shown that the system can perform nonlinear regression and synthesis tasks by exploiting the high-dimensional, nonlinear embedding induced by its quantum dynamics. For temporal prediction, the absence of memory imposes a fundamental constraint—yet this can be systematically overcome through post-processing delay embeddings. These serve not only to enhance memory, but can effectively \textit{replace} it, enabling an otherwise minimal reservoir to perform temporal tasks. By effectively lifting the system into a higher-dimensional space, these embeddings reconstruct the attractor geometry underlying the target dynamics. Notably, the sharp transitions observed in VPTS as a function of $\delta$ reveal critical thresholds for achieving long-term stability, underscoring the dimensional nature of predictive capacity. 

While the present study does not directly benchmark quantum against classical reservoirs, it provides a principled baseline for future work exploring quantum advantage. The need for such investigation is underscored by emerging work questioning the standard narrative of quantum-enabled performance enhancements \cite{bowles_better_2024}. In particular, the observation that dynamical complexity — rather than quantum correlations \textit{per se} — underpins computational capacity raises important questions about when and where genuine quantum advantage emerges. Our results suggest that the key distinguishing feature of quantum reservoirs may be compressive: they can generate rich, high-dimensional dynamics with fewer physical resources, though realizing this advantage in practice remains a challenge for near-term devices. It is in the spirit of addressing these challenges that the presented scheme minimizes state preparation and control requirements. Consequently, it is straightforwardly implementable in a variety of current experimental platforms, including transmons and defect qubits. 

Overall, this minimal architecture offers a tractable testbed for probing the interplay between physical dynamics, effective dimensionality, and computational power. This opens new pathways for lightweight quantum machine learning, while providing a platform for clarifying the relationship between physical complexity, computational capacity, and effective dimensionality — an important target for future research. 

More broadly, this work echoes the trend underpinning many of the most \textit{transformative}\cite{vaswani_attention_2017} recent developments in machine learning - where improvements in model expressivity, versatility, and controllability often arise not from elaboration, but reduction. Identifying the minimal structures needed to manifest these properties means redundant architecture may then be excised. Computational power may ultimately be a function solely of dimensionality, but when designing the machinery to realise such spaces, less is more.

 \begin{acknowledgments}
This work was supported by the the European Union and the European Innovation Council through the Horizon Europe projects QRC-4-ESP (Grant Agreement no. 101129663) and QUEST (Grant Agreement No. 101156088), the UK Research and Innovation (UKRI) Horizon Europe guarantee scheme for the projects QRC-4-ESP  (Grant No. 101129663) and QUEST (Grant No. 10130220), and the UK Engineering and Physical Sciences Research Council (EPSRC, Grant No. EP/W028344/1). We acknowledge the use of the Lovelace HPC service at Loughborough University. 
\end{acknowledgments}

\section*{Author Declarations}
{\bf Conflict of Interest}\\
The authors have no conflicts to disclose.

\section*{Authors Contribution}
GM and JSTG made an equal contribution to the project. JSTG, SS, AZ, and AGB designed the research and supervised the project. GM, JSTG, and WO performed the main numerical calculations. GM led the development of the theoretical part and wrote the first draft of the manuscript. GM, JSTG, WO, SS, AZ, and AGB analyzed the data, interpreted the results, and contributed to writing and editing the paper. JSTG and AGB supervised the development of numerical models and codes.

\section*{Data Availability}
This research did not involve the analysis of any experimental data. The code used in numerical simulations is available from the corresponding authors upon reasonable request.

\section*{References}
% Create the reference section using BibTeX:
\bibliography{refs_submission}

\end{document}